\documentclass[doublecol]{epl2} 

\usepackage{xcolor}	
\usepackage{paralist}	%
\usepackage{empheq}
\usepackage{comment}
\usepackage{amsmath}
\usepackage{graphicx}%
\usepackage{bm}%
\usepackage{MnSymbol}
\usepackage{eufrak}
\usepackage{amsfonts}
\usepackage{import}
\usepackage{ulem}
\usepackage{float}
\usepackage[dvipsnames]{xcolor}

\usepackage{cuted}
\usepackage{capt-of}

\newcommand{\ket}[1]{| #1 \rangle}
\newcommand{\bra}[1]{\langle #1 |}

\newcommand{\slatticeA}{\mathrm{A}}
\newcommand{\slatticeB}{\mathrm{B}}

\newcommand{\SI}{Supplementary Material}

\definecolor{orange}{rgb}{1,0.5,0}

\title{Near-Field Mapping and Wavefunction Reconstruction of Topological Edge States in Molecular Aggregates}

\shorttitle{Near-field maps of topological  edge states of molecular aggregates} %

\author{A. Mishra\inst{1} %
\and S. Nayak\inst{1}\!\!\!\footnote{Present address: Institute of Physics,\unpenalty~Czech Academy of Sciences,\unpenalty~Cukrovarnick\'{a} 10,\unpenalty~Prague,\unpenalty~16200,\unpenalty~Czech Republic} \and A. Eisfeld\inst{1,2}\thanks{Email:\email{alexander.eisfeld@tu-dresden.de}}}
\shortauthor{A.~Mishra \etal}

\institute{                    
  \inst{1} Max Planck Institute for the Physics of Complex Systems, N\"{o}thnitzer Str.\,38, 01187 Dresden, Germany\\
  \inst{2} Institute of Theoretical Physics, TUD Dresden University of Technology, 01062 Dresden, Germany
}

\abstract{
Self-assembled two-dimensional molecular aggregates provide a versatile platform for engineering excitonic states, including topologically nontrivial edge states. 
However, these edge-states are difficult to probe optically: they have weak far-field transition strengths, and far-field spectra do not reveal whether an excitation is localized at the edge. 
For an aggregate consisting of two sublattices of molecules with genuinely complex transition dipoles, we show that spatially resolved near-field spectroscopy encodes the full spatial structure of these states. 
The resulting near-field maps do not mirror the site populations, but instead exhibit strongly nonuniform, interference-like features. 
We demonstrate that a convolutional neural network can reconstruct the complex expansion coefficients of the eigenstates from these maps, and that this reconstruction is robust against substantial random noise.
}

\begin{document}
\maketitle

\section{Introduction}

A paradigmatic manifestation of nontrivial topology in condensed-matter systems is the emergence of topological edge states, whose existence is understood through bulk–boundary correspondence \cite{Hatsugai1993,HasanKane2010,QiZhang2011}. 
For finite arrays of molecules coupled through long-range transition-dipole-dipole interactions, the effective Frenkel-exciton Hamiltonian can host in-gap edge-localized excitonic states of topological origin \cite{yuen2014topologically}. 
Such states are of interest because they can be spectrally isolated and comparatively robust against moderate disorder and structural imperfections, making them promising for controlled excitation dynamics in molecular architectures \cite{HasanKane2010,QiZhang2011}. 
In particular, Ref.~\cite{yuen2014topologically} predicted that such states can arise as delocalized excitations in two-dimensional molecular aggregates.

Two-dimensional molecular aggregates can self-assemble on metallic or dielectric surfaces \cite{SMaier_2008,hoffmann2017imaging,Mohnani2010, SCARFATO2008677, AMMON2022122097, goronzy2018supramolecular}. %
Specific molecular arrangements can be engineered through a suitable choice of substrate and molecule, for example, by modifying the side groups \cite{SMaier_2008,goronzy2018supramolecular,AMMON2022122097,subs_porphyrin_CMichael_langmuir}. 
Dielectric surfaces are often advantageous for probing exciton states because they typically preserve fluorescence, allowing one to infer the populations of different exciton states following optical or electronic excitation \cite{muller2011fluorescence, eisfeld2017superradiance, doppagne2017vibronic, luo2019electrically, maier2021molecular}. 
If the intermolecular separations are large enough that direct wavefunction overlap is negligible, the relevant coupling is governed by the molecular transition densities \cite{MayKuehn2011}. 
In many situations, this coupling is well approximated by the interaction between transition dipoles, an approximation that we also adopt here for simplicity. 
In this work, we consider the setup introduced in Ref.~\cite{yuen2014topologically}, which requires genuinely complex transition dipoles, i.e., transition dipoles whose complex phase cannot be removed by a global gauge choice. 
As shown there, molecules with degenerate transitions provide a suitable platform for realizing such dipoles.
Possible choices of molecules are metalloporphyrins  or metallophthalocyanines, which  possess two degenerate electronic states, $\ket{Q^{(x)}}$ and $\ket{Q^{(y)}}$, with transitions at optical frequencies\cite{birnbaum2014optical,SolomonLever2006,LeverGray1983IronPorphyrinsPartI,Gouterman1986Porphyrins,Vlaming_2014}.
With a magnetic field this degeneracy can be lifted and one obtains an effective simple model with one excited state per molecule that has a genuinely complex transition dipole which for certain molecular arrangements  exhibits excitonic topological edge states \cite{yuen2014topologically}.

\begin{figure}[btp]
\centering
     \includegraphics[width=7.cm]{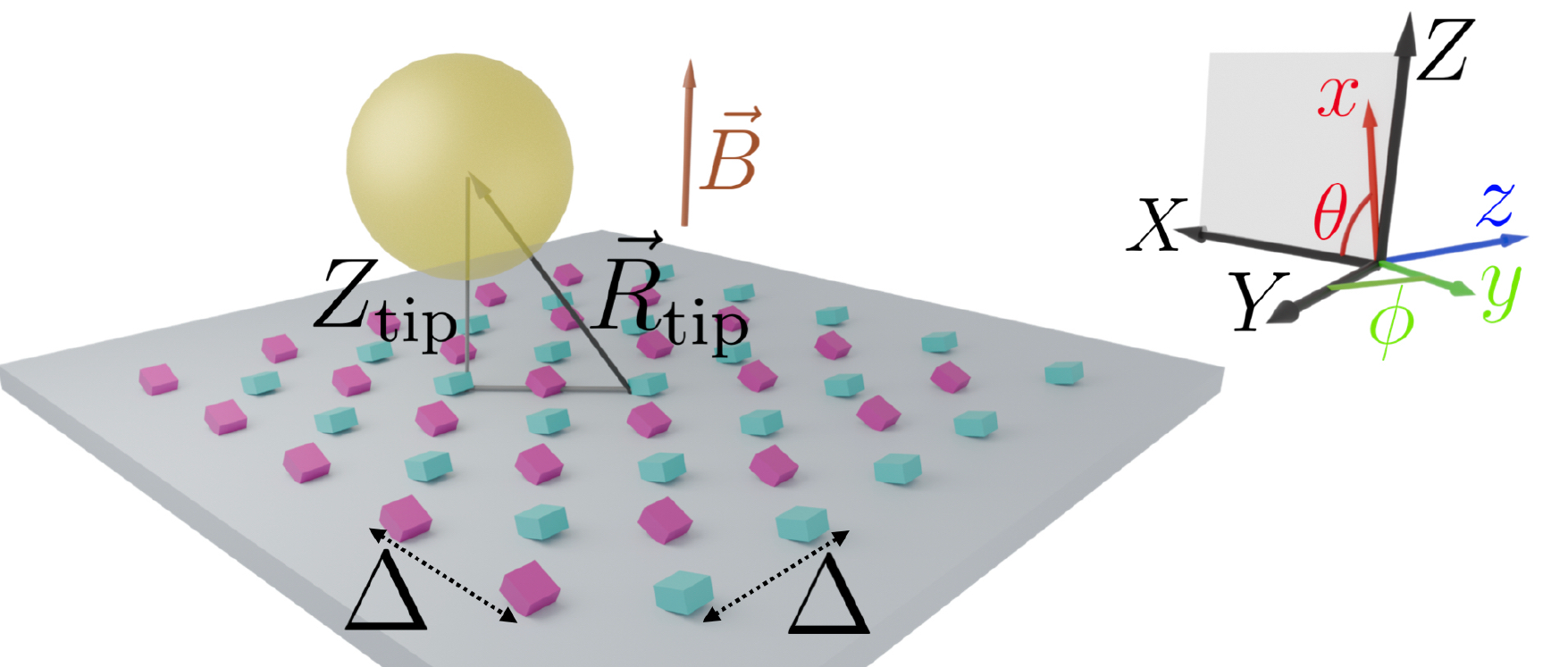}
\caption{ \label{fig:mol_lab_coordinate}
Schematic of the molecular aggregate and metallic tip. The tip apex is modeled as a spherical nanoparticle centered at $\vec{R}_{\mathrm{tip}} = (X_{\mathrm{tip}}, Y_{\mathrm{tip}}, Z_{\mathrm{tip}})$. 
A uniform magnetic field $\vec{B}$ is applied perpendicular to the aggregate plane. Neighboring molecules are separated by $\Delta$ within the same sublattice and by $\Delta/\sqrt{2}$ between different sublattices. The laboratory $(X,Y,Z)$ and molecular $(x,y,z)$ coordinate systems are indicated.
}
   
\end{figure}

Experimental investigation of these edge states is challenging. On the one hand, they typically exhibit weak transition strength in far-field spectroscopy and are therefore optically difficult to access. 
On the other hand, far-field spectroscopy does not directly reveal whether the corresponding excitation is localized at the edge.
A possibility to overcome these limitations is near-field spectroscopy \cite{Jain_Alivisatos_PNAS,Takuya_Hajime_APL,Berweger_Raschke_JPCL,Hermann_Gordon_AR}.

Here we show that near-field spectroscopy gives access to the spatial structure of the edge-state wavefunctions. 
Writing an eigenstate as
$
\ket{\phi^{(\ell)}}=\sum_{m=1}^{N} c_{m}^{(\ell)} \ket{m}
$,
the two-dimensional near-field maps, obtained by recording spectra at  different locations of the near-field, contain information about the complex site amplitudes \(c_{m}^{(\ell)}\), where \(\ket{m}\) denotes an excitation localized on molecule \(m\). 
However, the near-field maps of the edge states do not directly reflect the site populations \(|c_{m}^{(\ell)}|^{2}\) along the edge. 
Instead, they can display strongly nonuniform intensities, including weak signals at one or more edges and interference-like variations across the system, owing to the interplay between the complex coefficients \(c_{m}^{(\ell)}\) and the complex transition dipoles. Consequently, the near-field maps are difficult to interpret directly. 
Nevertheless, convolutional neural networks (CNNs) enable reconstruction of the coefficients \(c_{m}^{(\ell)}\), up to an overall phase, even when random noise is present in the near-field maps.
For small linear chains and small two-dimensional aggregates with real transition dipoles, this has been already achieved ~\cite{fulu2019ML}. 
We note that for various inverse problems in optical near-field microscopy  deep-learning methods have also successfully been applied \cite{Xu_Liu_acsnano,Zhao_Liu_JAP,Chen_Liu_AM}. 

\section{The considered system}

We specify the positions and orientations of the molecules in the laboratory frame with coordinates $(X,Y,Z)$. 
A sketch of the arrangement and the coordinate systems is shown in Fig.~\ref{fig:mol_lab_coordinate}. 
The molecules are arranged on a rectangular lattice in the $X$-$Y$ plane. The lattice consists of two sublattices, denoted by $\slatticeA$ and $\slatticeB$.
The molecules belonging to sublattice $\slatticeA$ are located at positions $(n_X,n_Y,0)\Delta$, where $n_X$ and $n_Y$ are integers running from $1$ to $N_X$ and $1$ to $N_Y$, respectively. 
Here, $\Delta$ is the distance between two neighboring molecules belonging to the same sublattice. 
The molecules belonging to sublattice $\slatticeB$ are located at positions $(n_X+0.5,n_Y+0.5,0)\Delta$.

We consider molecules with two degenerate excited states, $\ket{Q^{(x)}}$ and $\ket{Q^{(y)}}$, with transition energy $\hbar\omega_Q$ with respect to the  ground state $\ket{g}$. 
Assuming the corresponding wave functions to be real, we define the two orthogonal real transition dipoles
$\vec{\mu}^{(x)} \equiv \bra{Q^{(x)}} \hat{\vec{\mu}} \ket{g}$ and
$\vec{\mu}^{(y)} \equiv \bra{Q^{(y)}} \hat{\vec{\mu}} \ket{g}$,
which determine the $x$- and $y$-axes of the molecular coordinate frame. 
Following Ref.~\cite{yuen2014topologically}, we specify the molecular frame by first requiring that the $x$-axis lies in the $X$-$Z$ plane. 
The remaining degrees of freedom are then fixed by the angle $\theta$ between the molecular $x$-axis and the laboratory $X$-axis and the angle $\phi$ between the molecular $y$-axis and the laboratory $Y$-axis. 
The corresponding transition dipoles can then be written as\footnote{
Because of the degeneracy of $\ket{Q^{(x)}}$ and $\ket{Q^{(y)}}$, one may equally well choose any rotated basis within the degenerate subspace.
The corresponding transition dipoles are obtained from $\vec{\mu}^{(x)}$ and $\vec{\mu}^{(y)}$ by a rotation about the molecular $z$-axis. 
Such a rotation affects $\vec{\mu}^{\pm}$ only by a global phase,
which does not affect the final results.
}
\begin{equation}
\begin{split}
    \vec{\mu}^{(x)} &= \mu \, \big(\cos\theta,\, 0,\, \sin\theta\big), \\
    \vec{\mu}^{(y)} &= \mu \, \big(-\sin\phi \sin\theta,\, \cos\phi,\, \sin\phi \cos\theta\big).
\end{split}
\end{equation}
The molecular $z$-axis is 
$
    \vec{e}_z = (-\cos\phi \sin\theta,\,-\sin\phi,\,\cos\phi \cos\theta)
$.
In the following, we take all molecules within a given sublattice to have the same orientation, specified by the angles $\theta_\slatticeA$ and $\phi_\slatticeA$ for sublattice $\slatticeA$, and by $\theta_\slatticeB$ and $\phi_\slatticeB$ for sublattice $\slatticeB$.
An applied magnetic field $\vec{B}$ leads to a Zeeman splitting resulting in two nondegenerate states
$\ket{Q^{\pm}}=\frac{1}{\sqrt{2}}\big(\ket{Q^{(x)}} \pm i \ket{Q^{(y)}}\big)$
with complex transition dipoles
$\vec{\mu}^{\pm}= \frac{1}{\sqrt{2}}\big(\vec{\mu}^{(x)} \mp i \vec{\mu}^{(y)}\big)$.
The corresponding transition energies are approximately given by \cite{yuen2014topologically,malley1968zeeman}
\begin{equation}
\label{eq:zeeman_splitting}
\hbar \omega_Q \mp \varepsilon^{\mathrm{ZS}},
\qquad \text{with} \qquad
\varepsilon^{\mathrm{ZS}} \approx m_{\mathrm{l}} \mu_\mathrm{B}\, \vec{B}\cdot \vec{e}_z,
\end{equation}
where $\mu_\mathrm{B}$ is the Bohr magneton and $m_\mathrm{l}$ is the magnetic quantum number defined by
$\hat{L}_z \ket{Q^{\pm}} = \pm \hbar m_\mathrm{l} \ket{Q^{\pm}}$,
with $\hat{L}_z$ denoting the orbital angular momentum operator with respect to the molecular $z$-axis.
We take the magnetic field to be parallel to the laboratory $Z$-axis. 
In the molecular frame, this gives
$\vec{B}\cdot \vec{e}_z = B \cos\phi \cos\theta$.
As can be seen from Eq.~(\ref{eq:zeeman_splitting}), the Zeeman shift $\varepsilon^\mathrm{ZS} $ depends on the orientation of the molecule relative to the magnetic field. 
For 
$(\theta_\slatticeA,\phi_\slatticeA)=(-60^\circ,0^\circ)$ and
$(\theta_\slatticeB,\phi_\slatticeB)=(0^\circ,60^\circ)$,
 the Zeeman shift is the same on both sublattices,
$\varepsilon^{\mathrm{ZS}} = m_\mathrm{l}\mu_\mathrm{B} B \cos(60^\circ)$.
In this work, we focus on molecular orientations close to these values.

To further simplify the treatment, we retain only the lower-energy Zeeman state. 
This is justified when the Zeeman splitting is much larger than the intermolecular interaction. 
Typical values for substituted porphyrins, metalloporphyrins, or metallophthalocyanines  are around $m_\mathrm{l}=9$ \cite{malley1968zeeman}.
For this value of $m_\mathrm{l}$ and a magnetic field of $10\,\mathrm{T}$, the Zeeman splitting is about $80\,\mathrm{cm}^{-1}$, which is large compared to the transition dipole-dipole interaction between two porphyrin molecules. 
For example, for a transition dipole moment of $3\,\mathrm{D}$ and a separation of $2\,\mathrm{nm}$, the interaction is roughly $6\,\mathrm{cm}^{-1}$.

Since $\varepsilon^{\mathrm{ZS}}>0$ for $m_\mathrm{l}=9$, the lower-energy states are $\ket{Q^{+}_{\slatticeA}}$ and $\ket{Q^{+}_{\slatticeB}}$. 
The corresponding transition dipoles are
$\vec{\mu}^{+}_{\slatticeA}
=
\frac{1}{\sqrt{2}}
\big(\vec{\mu}^{(x)}_{\slatticeA}-i\vec{\mu}^{(y)}_{\slatticeA}\big)$
and
$\vec{\mu}^{+}_{\slatticeB}
=
\frac{1}{\sqrt{2}}
\big(\vec{\mu}^{(x)}_{\slatticeB}-i\vec{\mu}^{(y)}_{\slatticeB}\big)$,
which are genuinely complex. 
Restricting the description to the lower Zeeman branch also supports neglecting vibrational effects for low temperatures and weak intermolecular interactions; see, for example, the discussion in Ref.~\cite{eisfeld2017superradiance}. 
Throughout this work, we use $m_\mathrm{l}=9$.

As shown in Ref.~\cite{yuen2014topologically}, for the above choice of parameters, the Hamiltonian in Eq.~(\ref{eq:ham_topo}) exhibits eigenstates with probabilities $|c_m^{(\ell)}|^2 $ that are strongly localized at the edges. 
That work also showed that these edge states are of topological origin.
In the following, we investigate these states in more detail, with a particular focus on their near-field spectroscopic signatures.
\begin{figure*}[!tbp]
\centering
\includegraphics[width=14cm]{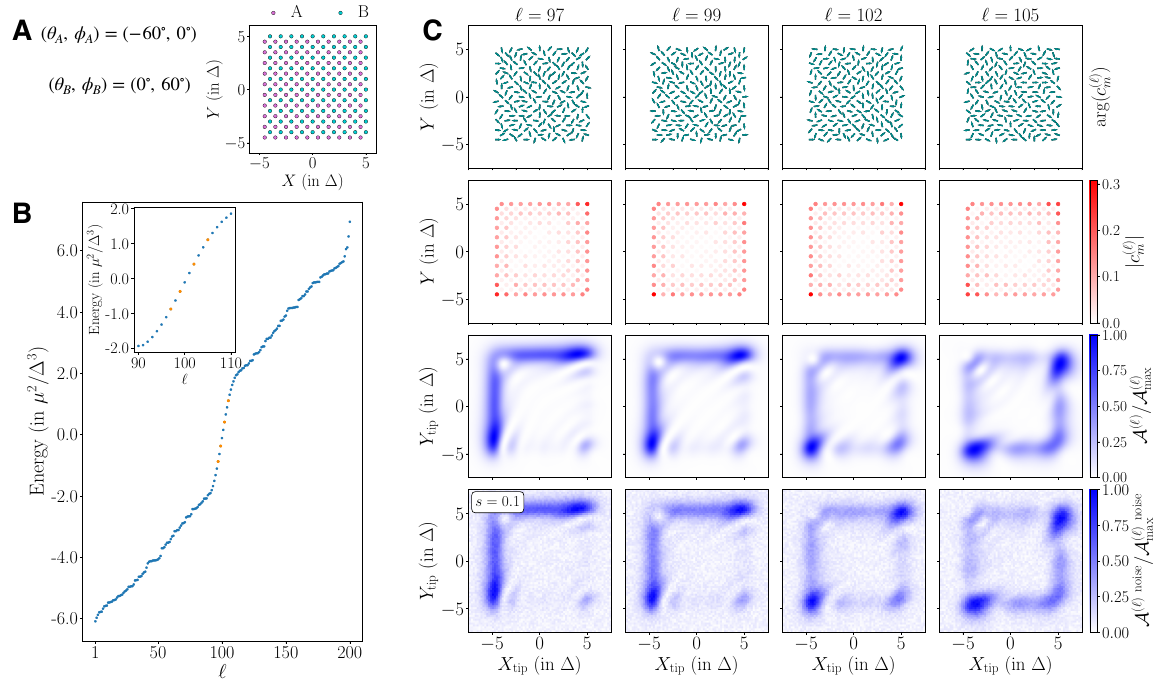}%
    \caption{ \label{fig:combined_sketch_energy_eigenstate_absorption}
Edge-localized eigenstates and corresponding near-field absorption maps for an aggregate with $N_X=N_Y=10$ sites on each sublattice and dipole orientations $(\theta_A,\phi_A)=(-60^\circ,0^\circ)$ and $(\theta_B,\phi_B)=(0^\circ,60^\circ)$. 
\textbf{(A)} Molecular arrangement of the two sublattices. 
The sublattices are distinguished by color. 
\textbf{(B)} Eigenenergies versus eigenstate index $\ell$, with the zero of energy set by the molecular on-site energy. 
The inset shows an enlarged view of the gap region.
\textbf{(C)} Four representative edge states (indicated in orange in (B)).
Top row: phases $\arg(c_m^{(\ell)})$, represented by unit arrows in the molecular plane. 
Second row: magnitudes $|c_m^{(\ell)}|$. 
Third row: normalized near-field absorption maps, $\mathcal{A}^{(\ell)}/\mathcal{A}^{(\ell)}_{\mathrm{max}}$, shown as a function of the tip position, where $\mathcal{A}^{(\ell)}_{\mathrm{max}}$ is the maximum value of the  $\mathcal{A}^{(\ell)}(\vec{R}_{\mathrm{tip}})$. 
Bottom row: corresponding normalized noisy absorption maps obtained by adding noise of strength $s=0.1$.  
Here $\mathcal{A}{}^{(\ell)~\mathrm{noise}}$ is the map obtained after the addition of noise, and its maximum absorption strength  $\mathcal{A}^{(\ell)~\mathrm{noise}}_{\mathrm{max}}$ is used for normalization.
The tip is scanned parallel to the molecular plane at a height of $2\Delta$, with its dipole oriented along the $Z$ axis.
Additional eigenstates are shown in the \SI.}
 \end{figure*}

\section{Theoretical modeling}

The $N$ molecules comprising the aggregate are placed at the positions $\vec{R}_n$ specified above. 
For each molecule $m$ we consider two electronic states, $\ket{g}_m$ and $\ket{e}_m$, corresponding to the ground and first excited state, respectively. 
The two states are separated by the energy $\varepsilon_m$, which is identical within each sublattice but can in general differ between the two sublattices, depending on the molecular orientation relative to the magnetic field. 
The associated transition dipole moments $\vec{\mu}_m$ are given by $\vec{\mu}^{+}_\slatticeA$ and $\vec{\mu}^{+}_\slatticeB$ for sublattice $\slatticeA$ and sublattice $\slatticeB$, respectively. 
Initially, the aggregate is in its global ground state $\ket{g_{\mathrm{agg}}}= \ket{g}_{1} \cdots \ket{g}_{N}$, in which all molecules are in their ground state. 
Since we are interested in linear spectroscopy, we focus on the case in which a single excitation is present in the aggregate. 
The relevant basis states are $\ket{m} = \ket{e}_m\prod_{n\neq m}^{N}\ket{g}_n$, where molecule $m$ is excited while all other molecules are in their ground states. 
The Hamiltonian of the aggregate in the subspace of these states is given by
\begin{equation}\label{eq:ham_topo}
	\begin{aligned}
		\hat{H}_{\mathrm{ex}} 
= \sum_{m=1}^N \varepsilon_m\ket{m}\bra{m} 
+ \sum_{\substack{m,n=1 \\ m \neq n}}^N V_{mn} \ket{m}\bra{n},
	\end{aligned}
\end{equation}
where the transition energies introduced above enter on the diagonal and $V_{mn}$ is the interaction between the transition densities \cite{MayKuehn2011} of molecule $m$ and molecule $n$. 
It is often admissible to approximate this interaction by that between point transition dipoles, in which case
\begin{equation}
\label{eq:Dip-Dip}
V_{mn}=\frac{1}{R_{mn}^3}\left(\vec{\mu}_m\cdot \vec{\mu}^*_n - 3 \Big(\vec{\mu}_m \cdot \frac{\vec{R}_{mn}}{R_{mn}}\Big)\Big(\vec{\mu}^*_n \cdot \frac{\vec{R}_{mn}}{R_{mn}}\Big)\right),    
\end{equation}
where $\vec{R}_{mn}$ is the displacement vector from molecule $m$ to molecule $n$ with magnitude $R_{mn}$. 
For the arrangement considered in the present work, the point transition dipoles $\vec{\mu}_m$ take the values $\vec{\mu}^{+}_\slatticeA$ and $\vec{\mu}^{+}_\slatticeB$ for sublattice $\slatticeA$ and sublattice $\slatticeB$, respectively. 
From the time-independent Schr\"odinger equation,
\begin{equation}
\label{eq:SE}
\hat{H}_{\mathrm{ex}}\ket{\phi^{(\ell)}}=E_\ell\ket{\phi^{(\ell)}},
\end{equation}
we obtain the $N$ eigenenergies $E_\ell$ and the corresponding eigenstates
\begin{equation}
\label{eq:eig_state}
 \ket{\phi^{(\ell)}}=\sum_{m=1}^{N}c_{m}^{(\ell)}\ket{m}. 
 \end{equation}

 In modeling the near-field maps, we follow our previous work \cite{gao2018near,nayak2021near}. 
 The strength of the transition from $\ket{g_{\mathrm{agg}}}$ to an eigenstate $\ket{\phi^{(\ell)}}$ is proportional to the absorption strength $\mathcal{A}^{(\ell)}$, given by
\begin{equation}\label{eq:A_l}
	\mathcal{A}^{(\ell)}(\vec{R}_{\mathrm{tip}})= \left|\sum_{m=1}^{N}c_{m}^{(\ell)}\, \vec{\mu}^*_{m}\cdot\vec{E}_m(\vec{R}_{\mathrm{tip}})\right|^2.
\end{equation}	
Here, $\vec{E}_m(\vec{R}_{\mathrm{tip}})$ is the electric field, oscillating at the transition frequency $E_\ell/\hbar$, evaluated at the center of molecule $m$. 
The argument $\vec{R}_{\mathrm{tip}}$ indicates that this field depends on the location of the tip (cf.\,Fig.\,\ref{fig:mol_lab_coordinate}).

\section{Topological characterization}
In the following investigations, we  also consider orientations that are different from the one investigated in Ref.~\cite{yuen2014topologically}.
For many of these orientations, the two sublattices will no longer have the same energy and it is not a priori clear that topological edge states still exist.
To ensure that we focus on orientations for which one has topological edge states, we have first studied the Hamiltonian (\ref{eq:ham_topo}) with periodic boundary conditions to identify regions in orientation space, for which the Chern number is non-zero.
We then restricted our investigation of the finite-size system to this regime, where we further identify topological edge states by their localization of population around the edge and by their chiral edge currents.
Details of our procedure are provided in the \SI{} (section 3).

\section{Near-field maps of topological edge states}

In all following calculations, we take the electric field to be that of an oscillating point dipole $\vec{d}$ located at $\vec{R}_{\mathrm{tip}}$. 
At the small tip--sample distances relevant for near-field spectroscopy, the near-field (quasistatic) component of this field dominates and is given, at position $\vec{r}$, by
\begin{equation}\label{eq:elec.field}
	\vec{E}(\vec{r};\vec{R}_{\mathrm{tip}}) = \frac{3(\vec{r}\!-\!\vec{R}_{\mathrm{tip}})\big[(\vec{r}\!-\!\vec{R}_{\mathrm{tip}}) \cdot \vec{d}\big]}{|\vec{r}-\vec{R}_{\mathrm{tip}}|^{5}} - \frac{\vec{d}}{|\vec{r}-\vec{R}_{\mathrm{tip}}|^{3}},
\end{equation}
so that the field entering Eq.~(\ref{eq:A_l}) is $\vec{E}_m(\vec{R}_{\mathrm{tip}})=\vec{E}(\vec{R}_m;\vec{R}_{\mathrm{tip}})$. 
In the present work, we choose $\vec{d}$ along the $Z$-axis, i.e., perpendicular to the molecular plane (in which the molecular point dipoles are located); for a general discussion of the influence of the direction of $\vec{d}$, see Ref.~\cite{nayak2021near}.

We take the distance from the molecular plane  to the tip position to be $Z_{\mathrm{tip}} = 2\Delta$. 
Clearly, the radius of the tip has to be smaller than this distance. 
For example, for a typical sublattice spacing of $\Delta = 2\,\mathrm{nm}$, one has $Z_{\mathrm{tip}} = 4\,\mathrm{nm}$. 
While we keep $Z_{\mathrm{tip}}$ fixed, we vary the in-plane position of the tip and show the absorption strength, Eq.~(\ref{eq:A_l}), as a function of $X_{\mathrm{tip}}$ and $Y_{\mathrm{tip}}$.

We choose a moderately sized aggregate, large enough to clearly exhibit the edge states but still small enough to allow reasonably fast numerical calculations. 
Specifically, we consider square sublattices with $N_X = 10$ and $N_Y = 10$, so that the aggregate consists of $N = 200$ molecules. 
For other sizes and for rectangular lattices, we find the same qualitative features.

In Fig.\,\ref{fig:combined_sketch_energy_eigenstate_absorption}, we show the eigenenergies and selected eigenstates for the dipole orientations $(\theta_\slatticeA, \phi_\slatticeA) = (-60^{\circ}, 0^{\circ})$ and $(\theta_\slatticeB, \phi_\slatticeB) = (0^{\circ}, 60^{\circ})$. 
These parameters lie inside the regime where one finds topological edge states (see Ref.~\cite{yuen2014topologically} and section 3.3.1 of  the \SI{}).
The states are labeled in order of increasing eigenenergy. The eigenenergies are shown in Fig.~\ref{fig:combined_sketch_energy_eigenstate_absorption}B. 
Energies are given in units of $\mu^2/\Delta^3$, and the zero of energy is set by the molecular on-site energy, which for this choice of angles is the same for all molecules, as discussed above.

One can identify two regions of densely spaced eigenenergies, at $E_\ell \lesssim -1.65$ and $E_\ell \gtrsim 1.75$, which we refer to as the lower and upper exciton bands, respectively. 
Between these two bands there are a few energetically well-separated states. 
This region is shown in more detail in the inset. 
In the \SI{}, we present additional calculations for aggregates with larger $N$. 
These results show that the energy ranges of the two bands are nearly independent of $N$. 
Both bands have a width of roughly $4\,\mu^2/\Delta^3$, while the band gap extends approximately from $E \approx -1.65$ to $E \approx 1.75$, corresponding to a width of $\Delta E_\mathrm{gap} \approx 3.4\,\mu^2/\Delta^3$ (for a more refined estimation see section S2 of the \SI{}). 
We also find that the number of states inside the band gap is approximately $N_\mathrm{gap}\approx\tfrac{1}{\sqrt{2}}(N_X+N_Y)$ (see the \SI{} for details). 
The typical energy spacing between the in-gap states can therefore be estimated as
$
    \delta E \sim \Delta E_\mathrm{gap}/ N_\mathrm{gap}
$.
To spectrally address individual states, the linewidth of the excitation laser should be smaller than this spacing, which is easier to achieve for smaller aggregates.
For $\mu=4\,\mathrm{Debye}$ and $\Delta=2\,\mathrm{nm}$, this yields $\delta E \approx 34\,\mathrm{cm}^{-1}/N_\mathrm{gap}$ and for the aggregate considered here ($N_X=N_Y=10$) we have $\delta E \approx 2.5\,\mathrm{cm}^{-1}$.

Inspection of the eigenstates shows that the states with eigenenergies inside the band gap have amplitude predominantly at the edges of the aggregate, whereas states in the two bands are distributed over large parts of the system and are suppressed near the boundaries. 
Since our main interest is in these edge states, we show representative examples in Fig.~\ref{fig:combined_sketch_energy_eigenstate_absorption}C. 
Specifically, we display the states with $\ell=97$, $99$, $102$, and $105$, corresponding to the highlighted energies in Fig.~\ref{fig:combined_sketch_energy_eigenstate_absorption}B. 
All states within the band gap, together with selected examples of bulk states, are shown in the \SI{}.
The upper two rows of Fig.~\ref{fig:combined_sketch_energy_eigenstate_absorption}C visualize the phase (top row) and magnitude (middle row) of the coefficients $c_m^{(\ell)}$ in the expansion~(\ref{eq:eig_state}). 
At each molecular position, an arrow indicates the phase of the corresponding coefficient $c_m^{(\ell)}$, while the color encodes its magnitude $|c_m^{(\ell)}|$. 
Note that $|c_m^{(\ell)}|^2$ is the probability of finding the excitation on molecule $m$ when the aggregate is in the eigenstate $\ket{\phi^{(\ell)}}$. 
For all four states shown in Fig.~\ref{fig:combined_sketch_energy_eigenstate_absorption}C, the excitation is distributed along the boundary and decays rapidly toward the center of the aggregate. 
The same behavior is found for the other states inside the band gap and becomes even more pronounced for larger $N$ (see the \SI{}). 
Although for the present system size, $N_X=N_Y=10$, the distinction between edge and bulk states is not perfectly sharp, we will refer to the states with $\ell=94,\dots,108$ as in-gap states or edge states; see also the discussion in the \SI{}.

While the magnitudes $|c_m^{(\ell)}|$ of the edge states follow rather similar spatial patterns, their phases differ substantially from state to state. 
As follows from Eq.~(\ref{eq:A_l}), the absorption strength depends not only on the magnitudes of the coefficients but also crucially on their relative phases.
Accordingly, the spatially resolved near-field absorption maps shown in the bottom row of Fig.~\ref{fig:combined_sketch_energy_eigenstate_absorption}C differ markedly between the four states. 
In all cases the signal is strongest near the edges, but the relative intensity varies strongly from one edge to another. 
For example, for the states $\ell=97$ and $\ell=99$ the absorption is concentrated mainly above the left and upper edges, with only weak signal elsewhere. 
By contrast, for $\ell=102$ and $\ell=105$ the lower and right edges also show appreciable absorption, with particularly strong signal near the lower-left and upper-right corners. 
In addition, all four states exhibit interference-like patterns, such that even along the edges there are regions where the absorption is strongly suppressed.

At first sight, there is therefore no simple relation between the spatially resolved near-field absorption maps and the coefficients $c_m^{(\ell)}$. 
For some edge states, such as $\ell=97$, it is even difficult to infer from the absorption map that the excitation probability is distributed rather uniformly along the boundary. 
This is consistent with our earlier observation that reconstructing the coefficients $c_m^{(\ell)}$ from spatially resolved near-field data is highly nontrivial~\cite{fulu2019ML}. 
The reason is that, according to Eq.~(\ref{eq:A_l}), the near-field absorption maps and the wavefunction coefficients are related nonlinearly, which makes the inversion problem difficult. This inversion becomes even more difficult, when the resolution is decreased and when there is noise in the near-field maps.
The last row of Fig.~\ref{fig:combined_sketch_energy_eigenstate_absorption} shows the near-field maps from the row above with reduced resolution ($75\times 75$ tip positions) and noise at each point sampled uniformly between $0$ and $s$ times the maximum of the respective map without noise (here $s=0.1$). 
One sees that small features, which could be used to reconstruct the coefficients $c_m^{(\ell)}$ are washed out.
It is therefore not clear a priori whether the near-field data contain sufficient information to reconstruct the underlying wavefunction.

In Ref.~\cite{fulu2019ML}, it was shown that such an inversion is possible for small linear chains and small two-dimensional aggregates with real transition dipoles by using a convolutional neural network (CNN). 
It was even possible to recover the coefficients $c_m^{(\ell)}$ from near-field maps to which a moderate amount of noise had been added. 
In the following, we show that a similar reconstruction is also possible for the edge states considered here. 
Compared with the previous work, the present problem is considerably more challenging: the number of molecules is much larger, and the coefficients are now complex. 
For example, for the aggregate studied in the present work, $200$ complex coefficients have to be determined. 
Furthermore, the near-field absorption maps of different edge states can look rather similar; see, for example, the cases $\ell=97$ and $\ell=99$, as well as the additional near-field maps shown in the \SI{}. 
Especially in the presence of noise, it is therefore not obvious whether a neural network can learn the subtle features that distinguish these states and predict the corresponding coefficients with useful accuracy. 
As we show below, this is nevertheless possible.

\begin{figure}[tbp]
\centering
\includegraphics[width=7cm]{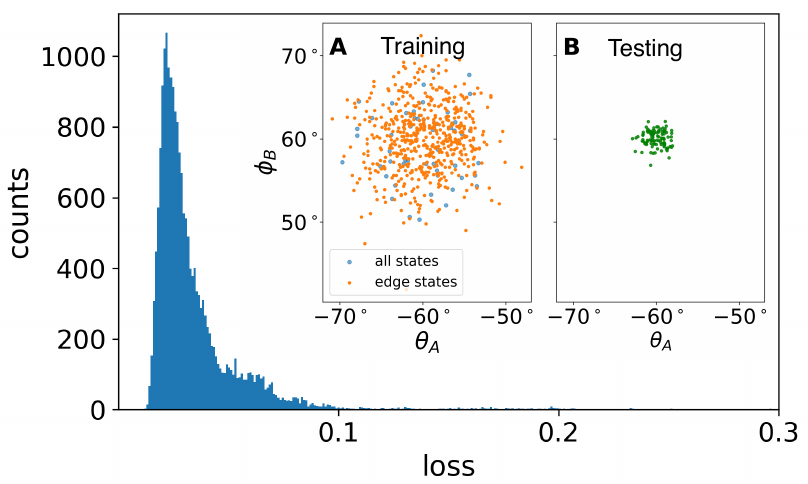} %
    \caption{
     Distribution of the loss values of the test dataset. 
    The distribution is computed using a histogram of 1000 bins.
    Insets: angle pairs $(\theta_A,\phi_B)$ used for (A) the training dataset and (B) the test dataset. Blue dots represent the angle pairs for which all eigenstates are considered, while the orange and green dots represent those for which only 20 states around the center of the band gap are included. 
     For both training and testing, noise with the same strength, $s=0.1$, was used.}
    \label{fig:Loss-distr}
\end{figure}

\begin{figure}[tbp]
    \centering
    \includegraphics[width=8.5cm]{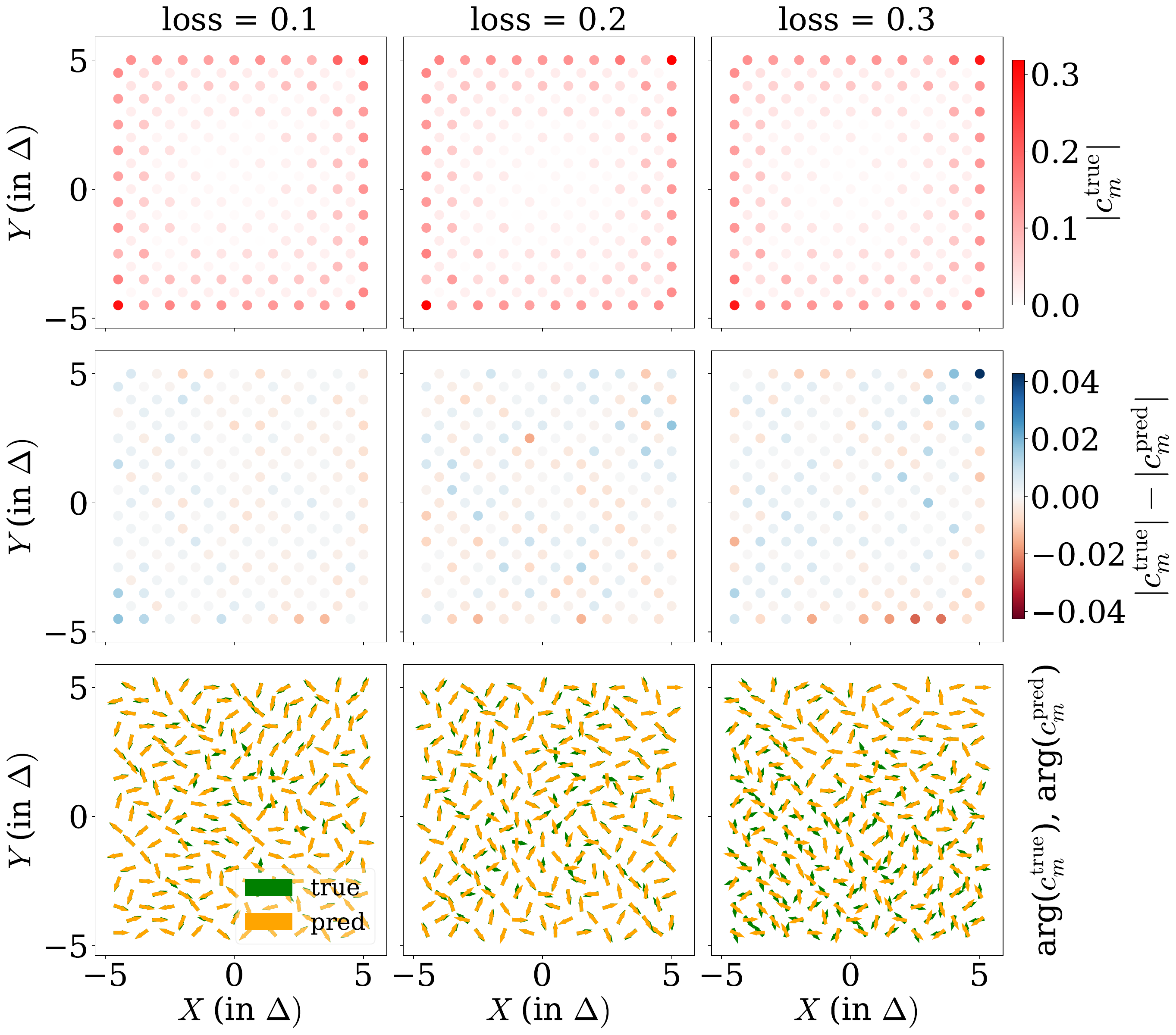} %
    \caption{
  Comparison between true and predicted excitonic edge-state coefficients for three states with different loss values.
  In each column, the top panel shows the amplitude of the true coefficients ($|c_m^{\text{true}}|$).
    The  middle panel shows the differences between the true and predicted amplitudes.
    The bottom panel compares the phases, with green arrows denoting the phases of the true coefficient ($\text{arg}(c_m^{\text{true}})$) and orange arrows denoting the phases of the predicted coefficients ($\text{arg}(c_m^{\text{pred}})$).
    The corresponding loss value is provided above each column.
    States with similar loss values show similar level of agreement between true and predicted coefficients. 
    \label{fig:loss_examples}}
    \end{figure}

\section{Predicting the wavefunction coefficients from the near-field maps}

Following Ref.~\cite{fulu2019ML}, we use a convolutional neural network (CNN) to reconstruct the eigenstates from the near-field maps. 
The inputs are near-field maps of $75\times 75$ data points (for examples see the bottom row of Fig.~\ref{fig:combined_sketch_energy_eigenstate_absorption}C), and the outputs are the $200$ complex coefficients $c_m^{(\ell)}$. 
The network architecture and training details are given in the \SI{} (section 4). 
As loss function we use the Frobenius norm\footnote{The Frobenius norm of a matrix $A$ is $\lVert A \rVert_\mathrm{F} = \sqrt{\sum_{m,n} |A_{mn}|^2}$.} of the difference between the true and predicted density matrices,
$L = \tfrac{1}{\sqrt{2}}\,\lVert \rho^{(\mathrm{true})} - \rho^{(\mathrm{pred})} \rVert_\mathrm{F}$,
where $\rho = \ket{\phi}\bra{\phi}$ and the eigenstate index $\ell$ has been suppressed. 
The prefactor $1/\sqrt{2}$ constrains $L$ to $[0,1]$ for normalized states.

Both training and testing are performed with random noise  added to the near-field maps (see the bottom row of Fig.\,\ref{fig:combined_sketch_energy_eigenstate_absorption}C and Fig.\,S11 of the \SI{} for typical noisy near-field maps). 
Training is carried out with small variations of the angles $\theta_\slatticeA$ and $\phi_\slatticeB$ (the other two angles fixed at $0^\circ$), which keeps the system in the topological regime (see in particular Fig.\,S9 of the \SI{}); the sampled angle pairs are shown in inset A of Fig.\,\ref{fig:Loss-distr}. 
To obtain a balanced training set with comparable numbers of edge and bulk states, part of the data includes all $200$ eigenstates and part includes only 20 states ($\ell=91$ to $\ell=110$) around the center of the band gap (which contain mostly topological edge states), for a total of about $500\,000$  near-field maps for training (20\% of these samples are randomly selected to form the validation dataset; the remaining 80\% are the training dataset). 
Testing is performed for 100 angle pairs, sampled independently of the training and validation data, from a small region (cf.~Fig.~3B) around the reference orientations $(\theta_A,\phi_A)=(-60^\circ,0^\circ)$ and $(\theta_B,\phi_B)=(0^\circ,60^\circ)$, using 20 states  around the center of the band gap; to each of the resulting 2\,000 near-field maps ten noise realizations have been added (in the following results the same noise strength $s$ as for the training has been used), resulting in a total of 20\,000 samples.
More details about training and testing are provided in the section 4 of the \SI, where also testing data outside the Hamiltonians used for training is considered (e.g., we added disorder on the site energies and all orientations).  

The main panel of Fig.\,\ref{fig:Loss-distr} shows that the resulting loss distribution peaks around 0.02, with nearly all values falling below 0.1.
To contextualize these metrics, Fig.\,\ref{fig:loss_examples} illustrates representative cases where $\text{loss} = 0.1$, $0.2$, and $0.3$.
We find that both magnitudes and phases are accurately reproduced for losses below 0.1, and even values around $0.2$ yield reasonable agreement. 
Because the loss distribution in Fig.\,\ref{fig:Loss-distr} lies almost entirely below 0.1, our model consistently achieves accurate reconstructions.

We note that, for simplicity, our training datasets were generated using a minimalist approach by varying only the angles $\theta_\slatticeA$ and $\phi_\slatticeB$.
Furthermore, we used a relatively small network and did not spend much time on optimization of its architecture.
Expanding the network, increasing the training time, and incorporating more diverse data (e.g., from a broader range of Hamiltonians) could further improve predictions, albeit at a rapidly growing computational cost. 
Furthermore, it remains unclear whether noisy near-field maps are distinct enough to ever permit perfect coefficient reconstruction. Nevertheless, the present results are sufficient to establish our central point: even from noisy near-field maps that carry appreciable absorption strength on only some edges, the CNN reconstructs the full eigenstate and clearly reveals its edge-state character.

\section{Conclusion}

We have investigated the near-field spectroscopic signatures of excitonic topological edge states in two-dimensional molecular aggregates. 
As a concrete platform, we considered aggregates of molecules with degenerate optical transitions in which an applied magnetic field lifts the degeneracy and gives rise to genuinely complex transition dipoles. 
For suitable molecular orientations, the resulting excitonic Hamiltonian supports eigenstates whose amplitudes are strongly localized at the edges of the aggregate.

Our central observation is that the near-field absorption maps of these edge states do not simply reflect the site populations $|c_m^{(\ell)}|^2$. 
Because of the interplay between the expansion coefficients $c_m^{(\ell)}$ and the transition dipoles, the maps exhibit strongly nonuniform intensities and interference-like features; in some cases, they show appreciable absorption only along part of the edge. 
As a consequence, the near-field maps cannot be interpreted directly as images of the excitation density. To overcome this difficulty, we employed a CNN to reconstruct the complex expansion coefficients of the eigenstates from the near-field maps. Importantly, this reconstruction remains reliable even when substantial random noise is added to the near-field maps during both training and testing.

Several methodological and physical extensions warrant further attention. On the methodological side, larger networks, more extensive training data, and longer training times are expected to further improve the reconstruction. 
It would also be useful to examine different types and strengths of noise. On the physical side, we expect the reconstruction of the eigenstate coefficients to improve when near-field maps for multiple excitation polarizations are taken into account, since different polarization directions can provide complementary information \cite{nayak2021near}. 
Incorporating vibronic states may likewise improve the reconstruction \cite{direct_wavefunction_YL_et_al}.
Additional natural extensions include going beyond the single-excitation regime, incorporating vibrational and finite-temperature effects, and applying the method to other lattice geometries and other classes of topological excitonic states. 
It would also be important to assess the influence of the tip on the eigenstates and the possibility of reconstructing them from the resulting near-field maps.

The use of a strong magnetic field may be experimentally inconvenient. 
In the present work, we employed it mainly to remain close to Ref.~\cite{yuen2014topologically} and to obtain a simple Hamiltonian with a single effective excited state per molecule. 
In principle, however, one could work without or with a small magnetic field by treating the two degenerate transitions directly, as done, for example, in Refs.~\cite{dolezal2021real,Vlaming_2014,YWan_2014,MARKOVITSI2001147}. 
In particular, the appendix of Ref.~\cite{Vlaming_2014} shows that, for a planar linear chain, two bands can arise naturally. 
For certain arrangements, one might even obtain topological edge states.
Recent investigations \cite{topo_noB_1,topo_noB_2} indicate that topological edge states can indeed be obtained without the need to induce  genuinely complex transition dipoles by a magnetic field. 
 Inducing a genuinely complex Hamiltonian, using  magnetic fields roughly an order of magnitude smaller than those used in Ref.~\cite{yuen2014topologically} and in the present work,  we have also found arrangements that exhibit topological edge states. 
 These small fields produce Zeeman splittings comparable to the interaction between neighboring molecules and are therefore not used to separate the two Zeeman components. 
 A possible drawback of such a scheme is that couplings to (low energy) vibrations may then become important. In all these investigations, one should keep in mind that the outermost row of molecules experiences a markedly different environment, with fewer neighboring molecules than those in the interior of the 2D domain, which may lead to different on-site energies for the edge molecules. If the difference in on-site energies between edge and interior molecules is large compared to the intermolecular interaction $V_{nm}$, trivial, non-topological edge states may arise.

Beyond optical near-field excitation, the edge states could also be probed using an STM tunnel current as the excitation source \cite{topo_noB_1,dolezal2021real,roslawska2026tip}.
Our reconstruction procedure can be adapted directly to this case.
It would also be interesting to investigate the applicability of time resolved nanoscopy \cite{Huber_Brixner_RevSciIns,VT_JCP}.
More broadly, the combination of spatially resolved near-field maps and machine-learning-based reconstruction may prove useful for characterizing a wide range of situations with topological states  in molecular nanostructures.

\acknowledgments
AE acknowledges support from the DFG via a Heisenberg fellowship (Grant No EI 872/10-1).


\begin{thebibliography}{10}
\expandafter\ifx\csname url\endcsname\relax\def\url#1{\texttt{#1}}\fi

\bibitem{Hatsugai1993}
\Name{Hatsugai Y.} \REVIEW{Phys. Rev. Lett.}{71}{1993}{3697}.

\bibitem{HasanKane2010}
\Name{Hasan M.~Z. \and Kane C.~L.} \REVIEW{Rev. Mod. Phys.}{82}{2010}{3045}.

\bibitem{QiZhang2011}
\Name{Qi X.-L. \and Zhang S.-C.} \REVIEW{Rev. Mod. Phys.}{83}{2011}{1057}.

\bibitem{yuen2014topologically}
\Name{Yuen-Zhou J. \etal} \REVIEW{Nat. Mater.}{13}{2014}{1026}.

\bibitem{SMaier_2008}
\Name{Maier S. \etal} \REVIEW{Small}{4}{2008}{1115}.

\bibitem{hoffmann2017imaging}
\Name{Hoffmann-Vogel R.} \REVIEW{Rep. Prog. Phys.}{81}{2018}{016501}.

\bibitem{Mohnani2010}
\Name{Mohnani S. \and Bonifazi D.} \REVIEW{Coord. Chem. Rev.}{254}{2010}{2342}.

\bibitem{SCARFATO2008677}
\Name{Scarfato A. \etal} \REVIEW{Surf. Sci.}{602}{2008}{677}.

\bibitem{AMMON2022122097}
\Name{Ammon M. \etal} \REVIEW{Surf. Sci.}{723}{2022}{122097}.

\bibitem{goronzy2018supramolecular}
\Name{Goronzy D.~P. \etal} \REVIEW{ACS Nano}{12}{2018}{7445}.

\bibitem{subs_porphyrin_CMichael_langmuir}
\Name{Milic T. \etal} \REVIEW{Langmuir}{20}{2004}{3974}.

\bibitem{muller2011fluorescence}
\Name{M{\"u}ller M. \etal} \REVIEW{Appl. Phys. B}{105}{2011}{67}.

\bibitem{eisfeld2017superradiance}
\Name{Eisfeld A. \etal} \REVIEW{Phys. Rev. Lett.}{119}{2017}{097402}.

\bibitem{doppagne2017vibronic}
\Name{Doppagne B. \etal} \REVIEW{Phys. Rev. Lett.}{118}{2017}{127401}.

\bibitem{luo2019electrically}
\Name{Luo Y. \etal} \REVIEW{Phys. Rev. Lett.}{122}{2019}{233901}.

\bibitem{maier2021molecular}
\Name{Maier S. \and St{\"o}hr M.} \REVIEW{Beilstein J.
  Nanotechnol.}{12}{2021}{950}.

\bibitem{MayKuehn2011}
\Name{May V. \and K{\"u}hn O.} \Book{Charge and Energy Transfer Dynamics in
  Molecular Systems} (Wiley) 2011.

\bibitem{birnbaum2014optical}
\Name{Birnbaum T. \etal} \REVIEW{J. Phys. Condens. Matter.}{26}{2014}{104201}.

\bibitem{SolomonLever2006}
\Name{Solomon E.~I. \and Lever A. B.~P.} (Editors) \Book{Inorganic Electronic
  Structure and Spectroscopy} Vol.~II (Wiley) 2006.

\bibitem{LeverGray1983IronPorphyrinsPartI}
\Name{Lever A. B.~P. \and Gray H.~B.} (Editors) \Book{Iron Porphyrins, Part
  {I}} (Addison-Wesley, Reading) 1983.

\bibitem{Gouterman1986Porphyrins}
\Name{Gouterman M., Rentzepis P.~M. \and Straub K.~D.} (Editors)
  \Book{Porphyrins: Excited States and Dynamics} Vol. 321 of \emph{ACS Symp.
  Ser.} (ACS, Washington) 1986.

\bibitem{Vlaming_2014}
\Name{Vlaming S.~M. \and Eisfeld A.} \REVIEW{J. Phys. D: Appl.
  Phys.}{47}{2014}{305301}.

\bibitem{Jain_Alivisatos_PNAS}
\Name{Jain P.~K. \etal} \REVIEW{Proc. Natl. Acad. Sci. U. S.
  A.}{109}{2012}{8016}.

\bibitem{Takuya_Hajime_APL}
\Name{Iida T. \etal} \REVIEW{Appl. Phys. Lett.}{98}{2011}{053108}.

\bibitem{Berweger_Raschke_JPCL}
\Name{Berweger S. \etal} \REVIEW{J. Phys. Chem. Lett.}{3}{2012}{945}.

\bibitem{Hermann_Gordon_AR}
\Name{Hermann R.~J. \and Gordon M.~J.} \REVIEW{Annu. Rev. Chem. Biomol.
  Eng}{9}{2018}{365}.

\bibitem{fulu2019ML}
\Name{Zheng F. \etal} \REVIEW{Phys. Rev. Lett.}{123}{2019}{163202}.

\bibitem{Xu_Liu_acsnano}
\Name{Xu S. \etal} \REVIEW{ACS Nano}{15}{2021}{18182}.

\bibitem{Zhao_Liu_JAP}
\Name{Zhao Y. \etal} \REVIEW{J. Appl. Phys.}{133}{2023}{133105}.

\bibitem{Chen_Liu_AM}
\Name{Chen X. \etal} \REVIEW{Adv. Mater.}{35}{2023}{2109171}.

\bibitem{malley1968zeeman}
\Name{Malley M. \etal} \REVIEW{J. Mol. Spectrosc.}{26}{1968}{320}.

\bibitem{gao2018near}
\Name{Gao X. \and Eisfeld A.} \REVIEW{J. Phys. Chem. Lett.}{9}{2018}{6003}.

\bibitem{nayak2021near}
\Name{Nayak S. \etal} \REVIEW{J. Chem. Phys.}{155}{2021}{134701}.

\bibitem{direct_wavefunction_YL_et_al}
\Name{Luo Y. \etal} \REVIEW{Sci. Adv.}{12}{2026}{eaeb6389}.

\bibitem{dolezal2021real}
\Name{Dole{\v z}al J. \etal} \REVIEW{ACS Nano}{16}{2022}{1082}.

\bibitem{YWan_2014}
\Name{Wan Y. \etal} \REVIEW{J. Phys. Chem. C}{118}{2014}{24854}.

\bibitem{MARKOVITSI2001147}
\Name{Markovitsi D. \etal} \REVIEW{Chem. Phys.}{269}{2001}{147}.

\bibitem{topo_noB_1}
\Name{Stein Q. \etal} arXiv:2610.09149 (2026).

\bibitem{topo_noB_2}
\Name{Nayak S. \etal} arXiv:2610.09148 (2026).

\bibitem{roslawska2026tip}
\Name{Ros{\l}awska A. \etal} \REVIEW{Rev. Mod. Phys.}{98}{2026}{025007}.

\bibitem{Huber_Brixner_RevSciIns}
\Name{Huber B. \etal} \REVIEW{Rev. Sci. Instrum.}{90}{2019}{113103}.

\bibitem{VT_JCP}
\Name{Tiwari V.} \REVIEW{J. Chem. Phys.}{154}{2021}{230901}.

\end{thebibliography}
\end{document}